%% file: main.tex
\documentclass[10pt,twocolumn,letterpaper]{article}

\usepackage{cvpr}              

\input{preamble}

\definecolor{cvprblue}{rgb}{0.21,0.49,0.74}
\definecolor{tabfirst}{rgb}{1, 0.7, 0.7} 
\definecolor{tabsecond}{rgb}{1, 0.85, 0.7} 
\definecolor{tabthird}{rgb}{1, 1, 1} 

\DeclareMathOperator*{\flip}{flip}
\usepackage{colortbl}
\usepackage[pagebackref,breaklinks,colorlinks,citecolor=cvprblue]{hyperref}

\def\paperID{16193} 
\def\confName{CVPR}
\def\confYear{2024}

\title{Passive Snapshot Coded Aperture Dual-Pixel RGB-D Imaging}

\author{Bhargav Ghanekar\\
Rice University\\
Houston TX USA\\
\and
Salman Siddique Khan\\
Rice University\\
Houston TX USA\\
\and
Pranav Sharma\\
IIT Madras\\
Chennai India\\
\and
Shreyas Singh\\
IIT Madras\\
Chennai India\\
\and
Vivek Boominathan\\
Rice University\\
Houston TX USA\\
\and 
Kaushik Mitra\\
IIT Madras\\
Chennai India\\
\and 
Ashok Veeraraghavan\\
Rice University\\
Houston TX USA
}

\begin{document}
\maketitle
\input{sec/0_abstract}    
\input{sec/1_intro}
\input{sec/2_relatedwork}
\input{sec/3_methods}
\input{sec/4_results}
\input{sec/5_applications}
\input{sec/6_conclusions} 
{
    \small
    \bibliographystyle{ieeenat_fullname}
    \bibliography{main}
}

\input{sec/X_suppl}

\end{document}

%% file: preamble.tex
\usepackage[dvipsnames]{xcolor}


%% file: sec/0_abstract.tex
\begin{abstract}
\vspace{-10pt}
Passive, compact, single-shot 3D sensing is useful in many application areas such as microscopy, medical imaging, surgical navigation, and autonomous driving where form factor, time, and power constraints can exist. Obtaining RGB-D scene information over a short imaging distance, in an ultra-compact form factor, and in a passive, snapshot manner is challenging. Dual-pixel (DP) sensors are a potential solution to achieve the same. DP sensors collect light rays from two different halves of the lens in two interleaved pixel arrays, thus capturing two slightly different views of the scene, like a stereo camera system. However, imaging with a DP sensor implies that the defocus blur size is directly proportional to the disparity seen between the views. This creates a trade-off between disparity estimation vs. deblurring accuracy. To improve this trade-off effect, we propose CADS (Coded Aperture Dual-Pixel Sensing), in which we use a coded aperture in the imaging lens along with a DP sensor. In our approach, we jointly learn an optimal coded pattern and the reconstruction algorithm in an end-to-end optimization setting. Our resulting CADS imaging system demonstrates improvement of $>$1.5~dB PSNR in all-in-focus (AIF) estimates and  5-6\% in depth estimation quality over naive DP sensing for a wide range of aperture settings. Furthermore, we build the proposed CADS prototypes for DSLR photography settings and in an endoscope and a dermoscope form factor. Our novel coded dual-pixel sensing approach demonstrates accurate RGB-D reconstruction results in simulations and real-world experiments in a passive, snapshot, and compact manner. 
\vspace{-15pt}
\end{abstract}

%% file: sec/1_intro.tex
\section{Introduction}
\label{sec:intro}
\begin{figure}[t!]
  \vspace{-10pt}
  \centering
  \includegraphics[width=\linewidth]{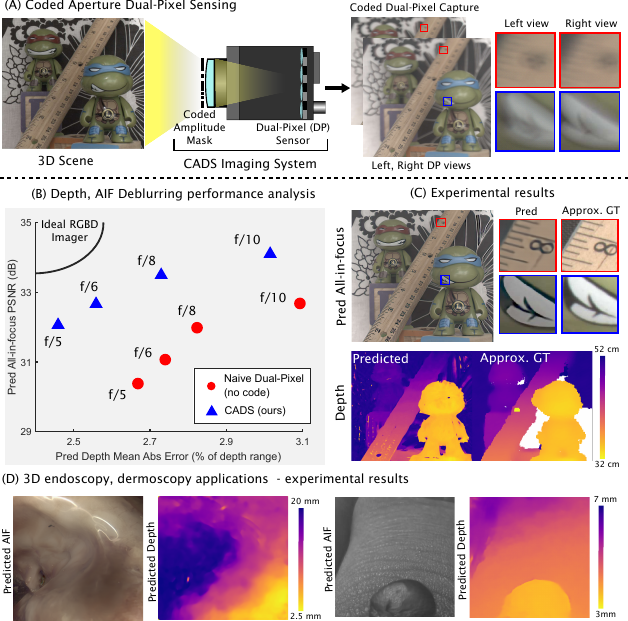}
  \caption{(A). We propose CADS - an imaging approach that leverages DP sensors and coded aperture masks for passive snapshot 3D imaging. (B) Coded DP sensing improves on naive dual-pixel sensing for simultaneous depth estimation and deblurring. Our CADS imaging prototype can recover accurate depth maps and all-in-focus images in (C) DSLR photography settings, as well as for (D) 3D endoscopy and dermoscopy. Read more on our website \url{https://shadowfax11.github.io/cads/}.}
  \label{fig:1}
  \vspace{-20pt}
\end{figure}
Extracting 3D information from 2D images has widespread applications in several domains such as microscopy, medical imaging, and autonomous driving. Conventional methods like passive stereo, structured illumination~\cite{boyer1987color,saxena2015structured}, photometric stereo~\cite{basri2007photometric}, and time-of-flight sensing do an excellent job estimating depth maps from 2D images for several computer vision tasks. However, these 3D imaging methods have a bulkier form factor due to the use of multiple cameras/illumination sources, and their estimation accuracy suffers when they are used in systems with size restrictions. 

Monocular depth estimation has the potential to overcome these form factor restrictions. A popular monocular depth sensing strategy is coded aperture imaging, in which the aperture of the lens is coded either using an amplitude~\cite{levin2007image, shedligeri2017data} or a phase mask~\cite{wu2019phasecam3d}. However, the quality of the depth maps and all-in-focus images obtained using these coded-aperture systems is still subpar compared to traditional methods like stereo or time-of-flight sensing. Recently, dual-pixel (DP) sensors have emerged as an accurate monocular depth sensing technology that overcomes these drawbacks due to their unique sensor design. These sensors are commonly found in smartphones and DSLR cameras. 

Although DP sensors were primarily developed for fast and accurate auto-focusing in consumer-grade cameras, there have been numerous works~\cite{pan2021dual,xin2021defocus,punnappurath2020modeling,Kim_2023_CVPR} that have tried to use these sensors as single-shot depth and intensity imaging sensors. Specifically, because of the unique micro-lens array arrangement over the photodiodes, these sensors can capture two blurry views of the same scene in a single capture. Moreover, the disparity between these two views is directly related to the defocus blur size. The small disparity in the views can be used for depth estimation while deblurring the defocus blur can provide us with the all-in-focus (AIF) image estimate. However, existing DP AIF estimation works are inherently limited by poor conditioning of the DP defocus blur. The inability to deblur with high fidelity then severely restricts their depth-of-field. 

\begin{figure*}[h!]
  \centering
  \includegraphics[width=\linewidth]{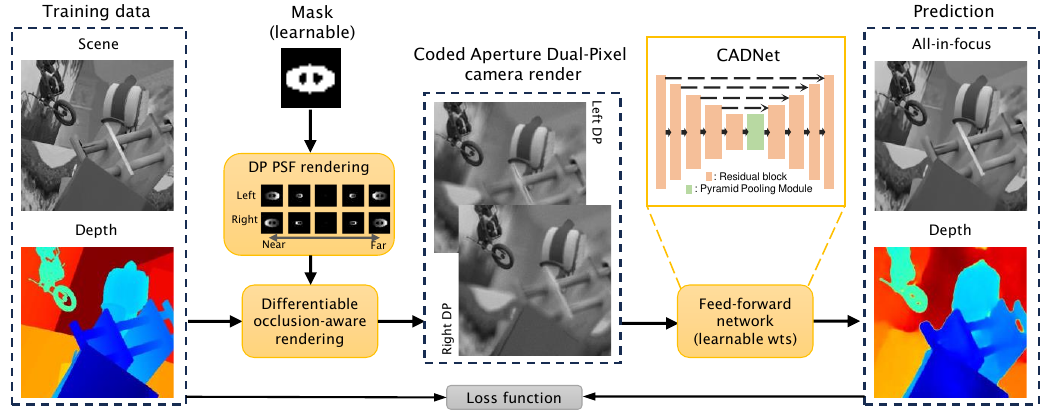}
  \caption{\textbf{Pipeline for Coded Aperture Dual-Pixel Sensing (CADS) and corresponding 3D scene estimation.} We perform end-to-end (E2E) optimization on simulated data, to learn an optimal amplitude mask and neural network weights for predicting deblurred all-in-focus (AIF) images and depth maps from coded dual-pixel captures. Our end-to-end learned system provides the best trade-off between depth estimation and AIF quality.}
  \label{fig:2-pipeline}
  \vspace{-15pt}
\end{figure*}

To improve the depth-of-field of dual-pixel cameras while still maintaining their ability to capture depth information, we introduce a novel sensing strategy called CADS - \textbf{C}oded \textbf{A}perture \textbf{D}ual-pixel \textbf{S}ensing. CADS uses a learned amplitude mask (code) in the aperture plane of a camera equipped with a DP sensor to improve the conditioning of the DP defocus blur while maintaining the disparity estimation ability from the two views. We use a two-stage, end-to-end learning framework to simultaneously learn the optimal coded mask pattern and the optimal neural network weights that can produce an accurate depth map and deblurred all-in-focus image estimates from dual-pixel measurements for a wide range of aperture settings (see Fig. \ref{fig:2-pipeline}). 


Compared to conventional DP sensing (with no coded aperture), the learned coded dual-pixel sensing system provides a better AIF performance, indicated through a consistent $\geq1.5$dB PSNR improvement in the quality of AIF images and a 5-6\% improvement in the quality of depth maps for a wide range of aperture settings, allowing a better disparity and AIF quality trade-off. We verify the efficacy of CADS through various simulations and real-world experiments. CADS provides a mechanism for high-fidelity depth and intensity imaging in a small form factor. We demonstrate its benefits in two relevant real-world applications - 3D endoscopy and extended depth-of-field (EDOF) dermoscopy. For both, we demonstrate promising real-world experiments using proof-of-concept prototypes built in the lab, see Fig. \ref{fig:1}. In summary, our contributions are: 
\begin{itemize}
		\item We propose coded aperture dual-pixel sensing, called CADS, that significantly improves the trade-off between quality of disparity and AIF estimates from dual-pixel sensors. 
		\item At the core of our method is the proposed differentiable coded dual-pixel sensing model that allows end-to-end learning of the coded aperture mask specific to  the dual-pixel sensor geometry. 
		\item We highlight the efficacy of our learned code design through various simulations and real-world experiments. 
        \item We show the effectiveness of coded aperture dual-pixel systems for close-range, compact RGB-D imaging by building a coded dual-pixel 3D endoscope and a coded dual-pixel dermoscope.  
\end{itemize}

%% file: sec/2_relatedwork.tex
\section{Related works}
\label{sec:related-work}
\subsection{Dual-Pixel sensing}
Dual-pixel (DP) sensors were initially used for fast auto-focus~\cite{sliwinski2013simple,jang2015sensor}, primarily using the fact that in-focus (out-of-focus) scene points show zero (non-zero) disparity. Since then, they have been used for several tasks. Exploiting the disparity cue, \cite{punnappurath2019reflection} demonstrated the use of DP sensors for reflection removal. \cite{wadhwa2018synthetic} used classical stereo algorithms for disparity estimation, to simulate shallow depth-of-field. Since then, there have been several works on disparity or depth estimation using dual-pixel sensors~\cite{Garg_2019_ICCV,punnappurath2020modeling,pan2021dual,xin2021defocus,Kim_2023_CVPR}. \cite{zhang20202} combines dual-pixel data with binocular stereo data to predict accurate disparity maps. Most of the methods use optimization techniques~\cite{punnappurath2020modeling,xin2021defocus} or leverage deep networks in supervised~\cite{Garg_2019_ICCV,pan2021dual,Kim_2023_CVPR} or self-supervised~\cite{Kim_2023_CVPR} settings. Apart from disparity estimation, the authors in \cite{kang2022facial} show normal estimation of human faces as well. \cite{xin2021defocus, pan2021dual} have been shown on unidirectional disparity only. \cite{Kim_2023_CVPR} addressed this and came up with a self-supervised method for DP images showing bidirectional disparity. There also have been numerous works on deblurring DP captures~\cite{abuolaim2020defocus,abuolaim2021learning,pan2021dual,xin2021defocus,yang2023k3dn}, which too are based on optimization methods or deep neural networks. In contrast to the above works, which mainly focus on post-processing conventional dual-pixel captures, we propose a novel modification to DP sensing strategy by adding a coded mask in the aperture plane that allows us to achieve better disparity and AIF quality trade-off for dual-pixel sensing.
\subsection{Coded aperture imaging}
Multiplexing of incoming light using a coded mask in the aperture plane has been used for numerous imaging applications over the years. Levin \etal \cite{levin2007image} used a designed amplitude mask in the aperture plane for estimating a depth map and an all-in-focus image from a single defocused image. Veeraraghavan \etal \cite{veeraraghavan2007dappled} used amplitude-coded aperture patterns for improved light-field imaging. Zhou \etal \cite{zhou2009coded} use a classical genetic optimization scheme to obtain coded aperture designs for estimating depth from coded defocus pairs. Since the design of the code or the mask pattern plays an important role in the downstream task, recently, there have been numerous works on using end-to-end learning schemes for designing the optimal coded mask patterns. Shedligeri \etal\cite{shedligeri2017data} learn the optimal amplitude mask pattern for estimating depth and all-in-focus image from a single defocused capture and report an improvement over the heuristically designed mask pattern proposed in \cite{levin2007image}. A phase mask can also be used to introduce coding in the aperture plane. Recently,  works like \cite{haim2018depth, wu2019phasecam3d} have used end-to-end learning to design the optimal phase mask profiles for improved 3D imaging. The proposed CADS system also uses an end-to-end learned amplitude mask in the aperture plane. Compared to existing coded aperture works, we use this amplitude mask for a DP sensor and are the first to do so. The use of a learned code in a DP setting allows the recovery of high-fidelity depth and AIF for a wide range of depth compared to standard coded-aperture imaging.

%% file: sec/3_methods.tex
\section{\textbf{C}oded \textbf{A}perture \textbf{D}ual-pixel \textbf{S}ensing (CADS)}
\subsection{Coded dual-pixel image formation model}
\begin{figure}[h!]
    \centering
    \includegraphics[width=\linewidth]{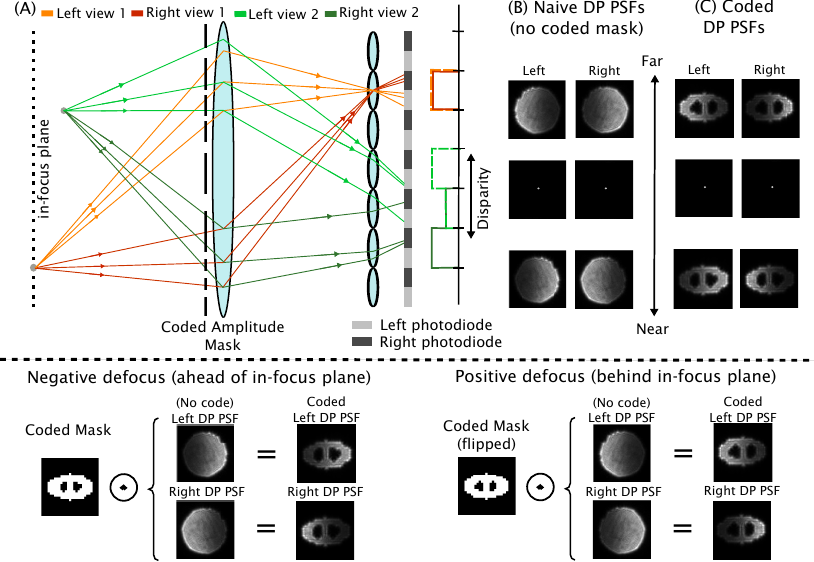}
    \caption{\textbf{CADS PSF formation model.} (A) With a DP sensor, scene points that are out-of-focus show disparity when comparing between left and right views. (B) This disparity is bi-directional, depending on scene depth relative to the in-focus plane. (C) When a coded amplitude mask is added to the imaging lens, the DP PSFs take the shape of the mask pattern. (D) Coded DP PSFs are a Hadamard product of the naive (no code) DP PSFs and the mask pattern.}
    \label{fig:dp-psf-formation}
    \vspace{-5pt}
\end{figure}

In a conventional image sensor, each pixel consists of one photodiode and a microlens to gather incoming light. In a dual-pixel (DP) sensor, each pixel consists of one microlens that covers two photodiodes, say left and right. Such a configuration enables all the left(right) photodiodes to receive light specifically from the right(left) half of the lens (see Fig.~\ref{fig:dp-psf-formation}A). Thus with a DP sensor, we obtain two images of the same scene from slightly different viewpoints, which are commonly referred to as the left and right DP images. Scene points that are in-focus are exactly aligned in both views, while scene points that are out-of-focus show positive or negative disparity when they are farther or nearer to the lens respectively. 

However, this disparity comes with an undesirable defocus blur effect of defocus blur. In a DP imaging system, the defocus blur size is directly proportional to the disparity. To have good, accurate depth measurements, one needs to have a large disparity range between the closest and the farthest scene depths, but this comes at the cost of highly defocused, blurry images, leading to poor AIF image estimation. 

To improve upon all-in-focus recovery while preserving the depth sensing capability of DP sensors, we propose the use of coded amplitude masks alongside DP sensors. 

When a coded (amplitude) mask $M$ is added to an imaging system/camera with DP sensor, the \textit{coded} DP PSFs can be expressed as a Hadamard (element-wise) multiplication of the naive DP PSFs with the coded mask. This is illustrated in Fig.~\ref{fig:dp-psf-formation}. More formally, the left and right DP PSFs ($h^{L,C}_z$ and $h^{R,C}_z$ respectively) can be expressed as 
\begin{equation}
    h^{L,C}_z = \flip_{z>z_{f}}(M) \odot h^{L}_z,  \;\;
    h^{R,C}_z = \flip_{z>z_{f}}(M) \odot h^{R}_z
\label{eq:coded-psfs-dualpix}
\end{equation}
where $h^{L}_z$, $h^{R}_z$ are the left and right view PSFs in the naive DP case (with no mask),  $M$ is simply the mask pattern (resized to the blur kernel size), $\flip_{z>z_{f}}(.)$ performs horizontal and vertical flipping when the point source is in positive defocus and is an identity function otherwise.

\subsection{CADS framework}\label{cads_framework}
Our framework for Coded Aperture DP Sensing (or CADS) consists of two components - 
\begin{itemize}
    \item \textbf{Coded dual-pixel image simulator}. Given scene intensity map (RGB or grayscale), scene depth map, coded mask pattern, and the naive (no code) DP PSFs, we use a differentiable rendering pipeline to generate the left, right CADS images of a scene. 
    \item \textbf{CADNet}. Given the left, right CADS images of a scene, we propose to use a neural network (which we call CADNet) that will take these images as inputs and produce a normalized defocus map and a deblurred all-in-focus (AIF) image as outputs.
\end{itemize}
We learn the two components in an end-to-end manner. 

\subsubsection{Coded dual-pixel image simulator}
\textbf{Mask-to-PSF generation}. In our CADS framework, we parameterize our mask pattern $M$ with a continuous-valued map $\theta_C(x,y)$ as done in \cite{shedligeri2017data}, where $M_{\theta_C}(x,y) = \sigma(\alpha* \theta_C(x,y))$, where $\sigma(.)$ is the sigmoid function. $\alpha$ is a scalar temperature factor that controls mask transparency for smooth learning\footnote{We gradually increase $\alpha$ over iterations to ensure the mask is learnt smoothly to be binary (0 or 1). See Supplementary for more details.}. For realistic modeling of the naive (no-code) DP PSFs $h^L_z$, $h^R_z$, we use the formulation provided in Abuolaim \textit{et al.}~\cite{abuolaim2021learning}. We generate left and right naive DP PSFs for $21$ depth planes, with signed blur sizes ranging from $-40$ pixels to $+40$ pixels. More details regarding this can be found in the supplementary. Given $M_{\theta_C}$, $h^L_z$, $h^R_z$, we use Eq.~\ref{eq:coded-psfs-dualpix} to generate the coded left, right DP PSF z-stack i.e. $h^{L,C}_z$ and $h^{R,C}_z$ for all the $21$ depth planes. 

\textbf{Coded DP image rendering.} We adopt a multi-plane image (MPI) representation of the scene, where the 3D scene is modelled as several discrete depth planes. Given $h^{L,C}_z$ and $h^{R,C}_z$, the coded DP left, right images $I_L$, $I_R$ of a 3D scene can be expressed as follows - 
\begin{equation}
\begin{split}
    I_L = \sum_z h^{L,C}_z * s_z, \;\;\;\; I_R = \sum_z h^{R,C}_z * s_z ,
\end{split}\label{eq:imaging-model-simple}
\end{equation}
where $s_z$ is the scene intensity at depth $z$, and $*$ is the 2D convolution operator. To remove artifacts at the edges of the MPI depth layers, we adopt the differentiable occlusion-aware modifications to Eqn.~\ref{eq:imaging-model-simple} from Ikoma \textit{et al.}~\cite{ikoma2021depth}, along with a minor modification. See supplementary for details. 

\subsubsection{CADNet for defocus and AIF estimation}
Recovering the depth map and the deblurred all-in-focus (AIF) image from blurry measurements can be solved by formulating a convex optimization problem~\cite{xin2021defocus}. However, such optimization methods are usually slow. Thus, we employ a neural network (referred to as CADNet), to recover depth and AIF images of the scene. CADNet architecture is based on the U-Net architecture~\cite{ronneberger2015u} that applies multi-resolution feature extraction with skip connections between encoder and decoder blocks of the same scale. We modify the basic U-Net architecture (illustrated in Fig.~\ref{fig:2-pipeline}) to have a residual block~\cite{he2016deep, zhang2018road} with two convolution layers of kernel size $3\times3$ at each scale. In addition, we append a pyramid pooling module (PPM)~\cite{zhao2017pyramid} at the lowest scale. A pixel-shuffle convolution layer~\cite{shi2016real} is used in the encoder for downsampling, while bilinear upsampling is used in the decoder. Our neural network is trained entirely on a synthetic dataset of coded DP images rendered from the FlyingThings3D dataset.  

CADNet takes as input the blurry measurements and outputs - (1) a normalized defocus map $\hat{D}_N$, ranging from -1 to +1, and (2) the deblurred all-in-focus (AIF) image $\hat{Y}$ of the scene. The normalized defocus map $\hat{D}_N$ is converted to actual physical defocus blur size by scaling it as $\hat{D} = \hat{D}_ND_{max}/p$, where $p$ is the pixel pitch of the DP sensor, and $D_{max}$ being the maximum blur size (in px) that CADNet is trained to handle. This defocus value is related to the imaging depth as follows~\cite{Garg_2019_ICCV} 
\begin{equation}
    \label{eq:coc-defocus}
    D(z) = \frac{Lf}{1 - f/g}\left(\frac{1}{g} - \frac{1}{z}\right), 
\end{equation}
where $L$ is the diameter of the imaging lens, $f$ is the focal length, $g$ is the in-focus distance, and $z$ is the depth of the object being imaged. Thus, given information about the imaging system we can readily calculate the depth from the CADNet defocus map output using Eq.~\ref{eq:coc-defocus}. For the AIF output channel from CADNet, we output normalized AIF images ranging from 0 to 1. We train two variants of CADNet, namely CADNet-Mono and CADNet-RGB. CADNet-Mono takes 2 input channels - the coded left, right DP images from a single, monochrome (grayscale) color, and the AIF outputs are single channel only. CADNet-RGB takes 6 input channels - the coded left, right DP images for each of the red, green, and blue channels, and the AIF output is a 3-channel RGB image. We train these two variants because real-world DP sensors present in the Pixel 4 and Canon EOS Mark IV DSLR are single-channel (green only) and full-frame (RGB) respectively. 

\textbf{Loss functions.} We jointly optimize for the mask pattern parameters ($\theta_C$) and CADNet weights ($\theta_D$) by minimizing the following loss function
\begin{align}
    \label{eq:loss}
    \mathcal{L}&=\mathcal{L}_{AIF}+\mathcal{L}_{defocus}+\mathcal{L}_{mask}, \\
    \mathcal{L}_{AIF}&=\beta_1\lVert\hat{Y}-Y_{gt}\rVert_1 + \beta_2\lVert\nabla\hat{Y}-\nabla Y_{gt}\rVert_1,\\
    \mathcal{L}_{defocus}&=\beta_3\lVert\hat{D}-D_{gt}\rVert_1 + \beta_4\lVert\nabla\hat{D}-\nabla D_{gt}\rVert_1,\\
    \mathcal{L}_{mask}&=\beta_5\text{Relu}\left(0.5 - \frac{\sum_{x,y}M_{\theta_C}}{\sum_{x,y}M_{open}}\right)
\end{align}
where  $\lVert.\rVert_1$ is the $L_1$ loss, $\nabla(.)$ is 2D gradient operator. $\hat{Y}$ and $\hat{D}$ are predicted AIF and defocus maps, while $Y_{gt}$ and $D_{gt}$ are groundtruth AIF and defocus maps. The last term is a mask pattern regularizer, where $M_{open}$ is the open aperture mask pattern. We set $(\beta_1,\beta_2,\beta_3,\beta_4,\beta_5)=(1,0.5,1,0.5,1,10^3)$, with $\beta_5>>1$ to enforce a strict transmission constraint ($\geq50\%$ light efficiency).
\subsubsection{Implementation details}
For training the CADS framework, we used simulated data generated using the forward model (discussed in Section \ref{cads_framework}) on  FlyingThings3D dataset scenes\cite{mayer2016large}. 20k scenes are split $80\%-20\%$ into training and validation sets. We rescale the disparity maps for each scene to convert them to depth maps having range $\mathcal{D}_T$ = [$32$~mm, $76$~mm]. During training of our coded DP simulator, camera parameters were set at $f=4.38$~mm, $L=f/1.73$, $g=45$~mm, ensuring that depth range $\mathcal{D}_T$ corresponded to blur sizes of $\leq40$ pixels (based on Eq.~\ref{eq:coc-defocus}). During testing on simulated data, we set camera parameters to match our real-world setting with focal length $f=50$~mm, aperture $L=f/4$, in-focus distance $g=40$~cm, with a depth range $\mathcal{D}_V$ = [$32$~cm, $52$~cm]. The signed blur sizes for $\mathcal{D}_V$ again corresponded to $\leq40$ pixels (pixel pitch, $p=10.72 \mu m$). Using the loss function described in Eqn.~\ref{eq:loss}, we perform end-to-end training, learning weights for CADNet and learning a coded mask pattern with light efficiency $>=50\%$. To account for light loss due to coded mask, we add appropriate heteroscedastic noise~\cite{foi2008practical} to simulated captures. We train the end-to-end framework using the Adam optimizer. We initialize the coded aperture mask with an open aperture pattern of size 21x21 pixels. 
More details are in the supplementary.

%% file: sec/4_results.tex
\section{Experiments and results}
\subsection{Comparison with naive standard and dual pixel sensing}
As a result of end-to-end training, we learn an optimal coded mask pattern that better conditions the coded DP blurry measurements, thus helping in good depth and AIF recovery. The learned mask and the corresponding left, right coded DP PSFs as a result of our end-to-end training have been depicted in Figs.~\ref{fig:2-pipeline},  \ref{fig:dp-psf-formation}. We refer the readers to the Supplementary for further illustrations and explanations. In this section, we show that our learned coded-aperture DP sensing is in fact an improvement over naive monocular standard and dual-pixel sensing in terms of both AIF and depth estimation quality. We use PSNR, SSIM, and LPIPS measures\cite{zhang2018unreasonable} for the evaluation of AIF quality. We use Root Mean Square Error (RMSE), Mean Absolute Error (MAE), and depth estimation accuracy within a threshold
margin ($\delta^1<1.05^1$) for the evaluation of depth prediction quality (see Supplementary for definitions).
\begin{table}[h!]
    \centering
    \begin{tabular}{c|ccc}
     \multicolumn{4}{c}{Depth Prediction} \\ \hline
     Imager & RMSE(mm)$\downarrow$ & MAE(mm)$\downarrow$ & $\delta^1$$\uparrow$ \\ \hline
     Naive SP & 48.51 & 29.37 & 0.688 \\
     Naive DP & \cellcolor{tabsecond}13.66 & \cellcolor{tabsecond}5.51 & \cellcolor{tabsecond}0.967 \\ 
     \textbf{CADS} & \cellcolor{tabfirst}12.57 & \cellcolor{tabfirst}5.15 & \cellcolor{tabfirst}0.972 \\ \hline \hline
     \multicolumn{4}{c}{AIF Prediction} \\ \hline
     Imager & PSNR(dB)$\uparrow$ & SSIM$\uparrow$  & LPIPS $\downarrow$\\ \hline
     Naive SP & 27.69 & 0.790 & 0.427\\
     Naive DP & \cellcolor{tabsecond}29.72 & \cellcolor{tabsecond}0.832 & 0.381\cellcolor{tabsecond} \\ 
     \textbf{CADS} & \cellcolor{tabfirst}31.20 & \cellcolor{tabfirst}0.865 & 0.337\cellcolor{tabfirst} \\ 
    \end{tabular}
    \caption{\textbf{Comparison with naive standard and DP - simulation.} Scenes with $20$~cm depth range, rendered with f/4 aperture. Our proposed CADS shows the best performance for joint depth and AIF recovery. Red highlights best, orange highlights second best. (SP: Standard-pixel, DP: Dual-pixel)}
    \label{tab:code-vs-nocode}
    \vspace{-20pt}
\end{table}
\begin{figure*}[!ht]
    \centering
    \includegraphics[width=0.95\linewidth]{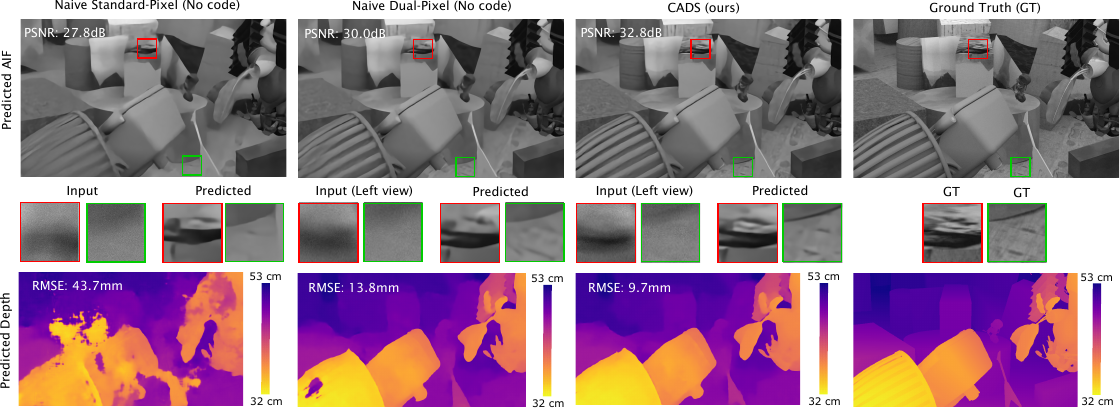}
    \caption{\textbf{Comparison with naive standard and DP - simulation.} DP sensing methods provide high-fidelity depth and all-in-focus (AIF) prediction compared to standard pixel sensing. Among DP sensing methods, CADS offers the best AIF and depth estimation quality observed through sharper AIF and cleaner depth.}
    \label{fig:sim-result}
    \vspace{-15pt}
\end{figure*}
\paragraph{Simulation results.}
We simulate measurements using the FlyingThings3D dataset and report the performance of CADS on it. We also compare CADS with two other imaging systems, namely, (a) Naive Std. Pixel - imaging system with a conventional standard pixel sensor and no aperture coding, (b) Naive Dual-pixel - imaging system with a dual-pixel sensor and no aperture coding. We trained a separate CADNet for each imaging system. Our simulation results are shown in Table~\ref{tab:code-vs-nocode} and Fig.~\ref{fig:sim-result}. When compared to the naive standard pixel scenario, both naive dual-pixel and CADS improve the deblurring and depth estimation accuracy significantly. Furthermore, CADS shows a $1.5$dB gain in PSNR and a $5$-$8$\% gain in depth prediction accuracy over naive dual-pixel. 

\paragraph{Real results.}
We perform real-world experiments by building a prototype imaging system capable of capturing coded dual-pixel images. We use the Canon EOS 5D Mark IV DSLR with a Yongnuo $50$~mm lens. The Canon EOS Mark IV sensor is an RGB, full-frame dual-pixel sensor, with every pixel in the R-G-G-B Bayer pattern being dual-pixels. Our coded mask pattern of diameter $L=f/4=12.5$~mm was printed on a transparency sheet and placed in the aperture plane of the Yongnuo lens. More details about the setup can be found in the supplementary.
\begin{figure*}[!ht]
  \centering
  \includegraphics[width=\linewidth]{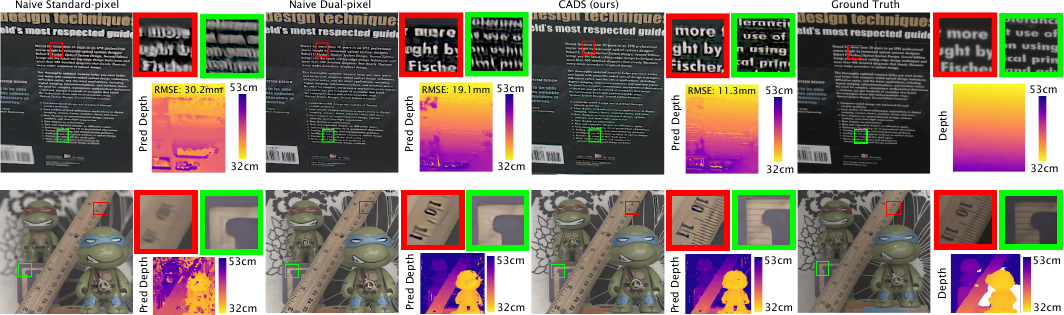}
  \caption{\textbf{Comparison with naive standard and DP - real data.} From left to right we show AIF and depth predictions by naive standard-pixel, naive dual-pixel and CADS, respectively, along with ground truth. CADS gives the best depth and AIF predictions. GT for AIF is obtained by capturing the scene with a $f/22$ aperture. A coarse GT depth map is obtained using the Intel RealSense sensor. Zoom in for best viewing.}
  \label{fig:macro-results-all-comparisons}
  \vspace{-10pt}
\end{figure*}
For accurate reconstruction, we render FlyingThings3D scenes with the real, captured PSFs, and fine-tune CADNet weights for 30 epochs with real-world captured PSFs. More details regarding fine-tuning procedure can be found in the supplementary. Fig.~\ref{fig:macro-results-all-comparisons} shows real-world results with our imaging prototype. Our proposed CADS system provides significantly better AIF estimates while preserving the depth-sensing ability of dual-pixel sensors. 

\begin{table}[!ht]
    \centering
    \begin{tabular}{c|cccc}
    \multicolumn{5}{c}{Depth Prediction MAE (mm) $\downarrow$} \\ \hline
     Imaging System & \multicolumn{4}{c}{Aperture size (f/\#)} \\ \hline 
      & f/4 & f/6 & f/8 & f/10 \\ \hline
Naive SP &   26.64 &                      26.45 &                      26.26 &                      24.39 \\
Coded SP &  \cellcolor{tabthird}21.37 &  \cellcolor{tabthird}22.18 &  \cellcolor{tabthird}21.83 &  \cellcolor{tabthird}22.68 \\
Naive DP   & \cellcolor{tabsecond}5.44 & \cellcolor{tabsecond}5.48 & \cellcolor{tabsecond}5.65 & \cellcolor{tabsecond}6.19 \\
\textbf{CADS} &  \cellcolor{tabfirst}4.99 &  \cellcolor{tabfirst}5.12 &  \cellcolor{tabfirst}5.46 &  \cellcolor{tabfirst}6.03 \\ \hline
\multicolumn{5}{c}{AIF Prediction PSNR (dB) $\uparrow$} \\ \hline
Imaging System & \multicolumn{4}{c}{Aperture size (f/\#)} \\ \hline 
      & f/4 & f/6 & f/8 & f/10 \\ \hline
Naive SP    &   27.66 &       29.16 &                      30.08 &                      30.87 \\
Coded SP &  \cellcolor{tabthird}28.97 &  \cellcolor{tabthird}30.28 &  \cellcolor{tabthird}31.16 &  \cellcolor{tabthird}31.90 \\
Naive DP   & \cellcolor{tabsecond}29.54 & \cellcolor{tabsecond}31.07 & \cellcolor{tabsecond}31.98 & \cellcolor{tabsecond}32.68 \\
\textbf{CADS}  &  \cellcolor{tabfirst}31.21 &  \cellcolor{tabfirst}32.66 &  \cellcolor{tabfirst}33.49 &  \cellcolor{tabfirst}34.10
    \end{tabular}
    \caption{\textbf{Performance under varying f-number}. The proposed CADS performs the best passive snapshot depth and all-in-focus estimation across different aperture sizes. Red highlights the best method, and orange highlights the second best.}
    \label{tab:comparison}
    \vspace{-15pt}
\end{table}
\subsection{Performance of CADS for various aperture sizes}\vspace{-1mm}
CADS offers a better trade-off between depth and AIF quality compared to a naive dual-pixel sensor for various aperture sizes as shown in Fig. \ref{fig:1}. To establish this, we evaluate the performance of CADS for various aperture sizes ranging from $f/4$ to $f/10$. We test this out on the same FlyingThings3D scenes. Apart from comparing CADS against naive dual-pixel sensing (Naive DP), we also compare it against coded and no-code variants of standard pixel sensor, referred to as Coded and Naive SP respectively. Table~\ref{tab:comparison} shows depth estimation and AIF deblurring results on our simulated FlyingThings3D data, where the imaging system parameters are the same as mentioned before, with objects being placed in the [32~cm, 53~cm] range. With decreasing aperture size (increasing f-number), AIF prediction accuracy increases while depth prediction gets worse. Our proposed E2E CADS system consistently shows the best performance across all apertures and improves the defocus-disparity trade-off.
\subsection{Comparison with existing dual-pixel works}\vspace{-1mm}
We compare the performance of CADS with existing learning-based dual-pixel sensing works, namely DPDNet\cite{abuolaim2021learning}, DDDNet\cite{pan2021dual}, Xin \etal \cite{xin2021defocus}, Punnapurath \etal \cite{punnappurath2020modeling} and Kim \etal \cite{Kim_2023_CVPR}. Since existing works were developed for naive (no-code) dual-pixel sensors, we use the naive DP PSF blur to simulate captures for evaluation. For evaluating CADS, we render using the coded-aperture DP PSF. We use a validation subset of 2k images from the FlyingThings3D dataset to evaluate. The results are shown in Table~\ref{tab:compare-with-previous}. Most existing works estimate disparity which is related to defocus map by an unknown scale. For evaluation, we use the affine invariant version of MAE (AI(1)) for disparity estimation quality~\cite{Garg_2019_ICCV} and PSNR for AIF quality\footnote{See Supplementary for metric definitions}. Note that we use the normalized defocus map output by CADNet for this comparison instead of converting it to a depth map. Works that were designed for unidirectional disparity~\cite{pan2021dual, xin2021defocus} show poor results when tested on simulated scenes with bidirectional disparity between the DP images. Our proposed CADS outperforms existing methods on the simulated FlyingThings3D dataset. Moreover, CADNet trained on naive DP blurs, referred to as Naive DP in Table~\ref{tab:compare-with-previous}, also outperforms existing methods. We also perform a comparison on rendered images from NYUv2 dataset\cite{silberman2012indoor} and report the performance in the supplementary.  
\begin{table}
    \vspace{0pt}
    \centering
    \begin{tabular}{c|c|c} 
    Method & AIF PSNR & Disp. AI(1)  \\ \hline
 DPDNet~\cite{abuolaim2021learning}& 24.34& N.A.\\ 
         DDDNet~\cite{pan2021dual}&21.35& 0.2769\\   
         Xin \etal ~\cite{xin2021defocus} & 18.13
      & 0.3018 \\   
 Punnappurath \etal ~\cite{punnappurath2020modeling} & N.A.& 0.1940\\  
 Kim \etal~\cite{Kim_2023_CVPR}& N.A.& 0.2866\\\hline
 Naive DP (ours)& \cellcolor{tabsecond}29.72& \cellcolor{tabsecond}0.0190\\
 \textbf{CADS (ours)}& \cellcolor{tabfirst}31.20& \cellcolor{tabfirst}0.0177\\ \end{tabular}
    \caption{\textbf{Comparison with existing DP methods.} CADS offers the best AIF and disparity estimation quality on our simulated dataset. Red highlights best, orange highlights second best.}
    \label{tab:compare-with-previous}
    \vspace{-15pt}
\end{table}

%% file: sec/5_applications.tex
\section{Applications}
We build two prototype CADS systems for 3D endoscopy and 3D dermoscopy and show real-world endoscopy and dermoscopy results using these prototypes. 
\subsection{CADS Endoscopy}
3D endoscopy is potentially useful for surgical navigation, tumor/polyp detection, etc~\cite{nomura2019comparison,bickerton2019three}. Stereo-endoscopes have been used for obtaining 3D information of the scene, but they are not compact. 3D endoscopy solutions have been attempted by using SfM methods \cite{ozyoruk2021endoslam}, but they lack accuracy due to non-rigid nature of human tissue and organs. Structured illumination methods have also been used for 3D endoscopy~\cite{furukawa2016shape}; however, this requires close control over illumination, which is difficult. Thus, there is a need for developing passive snap-shot depth sensing in endoscopy. 

\begin{figure}[h!]
    \centering
    \includegraphics[width=0.95\linewidth]{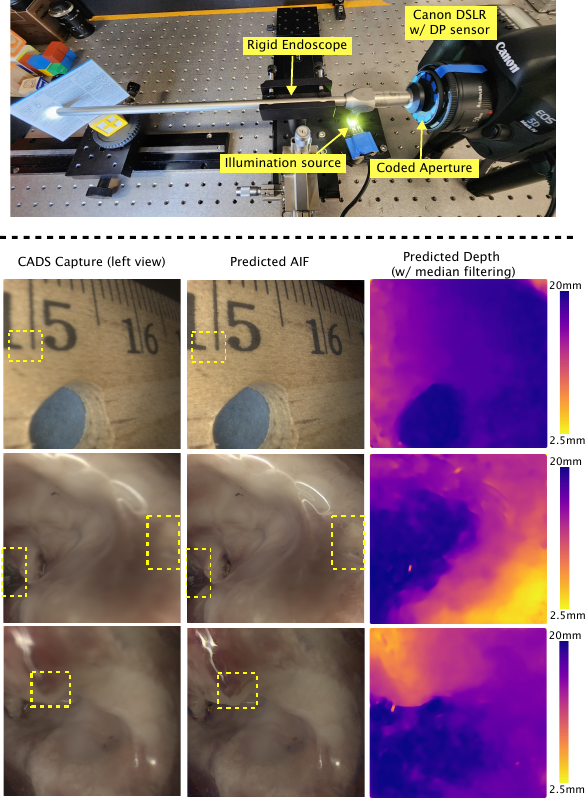}
    \caption{\textbf{CADS endoscopy results.} (Top) Proof-of-concept 3D endoscopy setup built using CADS. (Bottom) CADS allows us to recover sharp AIF and accurate depth for various scenes ($\sim$800x800 px recon.). Zoom-in on yellow boxes for best results.}
    \label{fig:endoscopy-results}
    \vspace{-15pt}
\end{figure}

CADS can potentially be a mechanism to achieve passive, snapshot, per-frame 3D endoscopy where depth maps as well as AIF images are simultaneously predicted. We built a coded aperture dual-pixel endoscopic sensing prototype as shown in Fig.~\ref{fig:endoscopy-results}. We use Canon EOS Mark IV DSLR with a Yongnuo $50$~mm lens as our dual-pixel camera and a Karl Storz Rubina 10mm Endoscope. The camera and the scope are coupled by aligning both their optical axis and placing the camera lens close to the endoscope eyepiece. We put our coded aperture pattern on a 3D-printed circular aperture and placed it in the middle of the scope and the DSLR lens. We focus our lens such that the in-focus plane is $2$~cm away from the other end of the endoscope. We capture PSFs for negative defocus and fine-tune only for the negative defocus region. To account for the spatial variance in the PSF, we randomly sample PSFs across the field of view for training. Details regarding the setup and PSF calibration are in the supplementary.

To emulate realistic endoscopy scenes, we image lamb chops tissue samples. Depth and AIF reconstruction results (image size $\sim$800x800) for such scenes are shown in Fig.~\ref{fig:endoscopy-results}. Depth maps are shown after 21x21 median filtering. Despite being finetuned on FlyingThings3D dataset, CADNet is still able to recover sharp AIF and clean depth maps. 

\subsection{CADS Dermoscopy}
\begin{figure}[h!]
    \centering
    \includegraphics[width=0.95\linewidth]{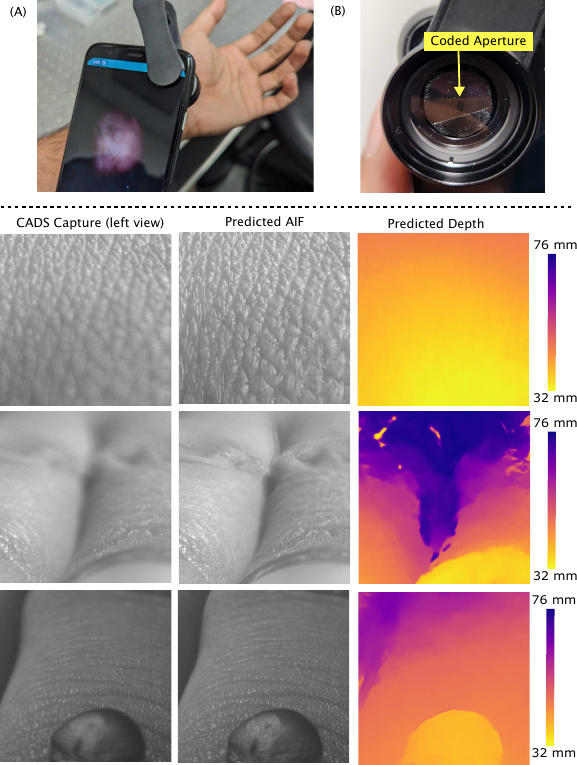}
    \caption{\textbf{CADS dermoscopy results.} (Top)(A)Prototype CADS dermoscope using Pixel 4 and $12\times$ macro lens.(B)The coded aperture is placed in between the smartphone lens and camera lens.(Bottom)Example AIF, depth reconstructions ($\sim$450x450 px)} 
    \label{fig:dermoscopy-result}
    \vspace{-20pt}
\end{figure}

\vspace{-2.5pt}
Dermoscopes are imaging systems used in clinical settings to view skin lesions. Existing non-contact dermoscopes suffer from small depth-of-field and can only provide 2D images. \cite{jutte2022focus} collect a focal stack to overcome this drawback. However, this is time-consuming, prone to motion blur, and doesn't provide depth information. We build an in-house smartphone-based CADS dermoscope,  using a Pixel 4 smartphone that is equipped with a DP sensor and 12x macro lens. We finetuned our CADNet model on measurements simulated from FlyingThings3D images using the captured spatially-varying PSFs. Fig. \ref{fig:dermoscopy-result} shows the visual results for depth and AIF predictions from our in-house dermoscope estimated using the fine-tuned CADNet on captures of a human subject's hand. The use of CADS in a dermoscope offers high-fidelity AIF and depth maps for a wide range of depth despite being finetuned on FlyingThings3D dataset.

%% file: sec/6_conclusions.tex
\vspace{-5pt}
\section{Conclusions}
\vspace{-2pt}
Dual-pixel sensors have emerged as a novel compact depth-sensing modality with widespread use in consumer-grade cameras. However, existing dual-pixel depth sensing methods suffer from limited depth of field (DOF) due to an inherent trade-off between the quality of depth maps and the ease of deblurring. Moreover, the majority of existing works fail for bidirectional defocus/disparity. To overcome these drawbacks, we use ideas from coded aperture imaging to develop a novel, single-shot 3D sensing strategy called CADS - Coded Aperture Dual-pixel Sensing. More specifically, we improve the conditioning of the dual-pixel defocus blur by introducing an end-to-end learned amplitude mask in the aperture plane. Introducing the learned mask improves the quality of deblurred all-in-focus images and depth maps for a wide range of depth. We demonstrate the efficacy of our proposed systems for three different applications: DSLR photography, 3D endoscopy, and extended DOF dermoscopy. Promising results on real data collected using proof-of-concept hardware indicate CADS can be used for applications beyond photography like medical imaging where form-factor plays a pivotal role. In the future, it would be interesting to replace amplitude masks with phase masks \cite{wu2019phasecam3d} that can improve the light efficiency and as a result, the SNR of the system.
\newline
\textbf{Acknowledgements.} This work was supported by NSF: IIS-1730574, NIH: R01DE032051-01. K.M. acknowledges NSF/IITM Pravartak Technologies Foundation and funding from the Department of Science and Technology, India. 

%% file: sec/X_suppl.tex
\clearpage
\setcounter{page}{1}
\maketitlesupplementary

\section{Methods}
\subsection{Realistic naive (no code) DP PSFs in simulations}
\begin{figure}[h!]
    \centering
    \includegraphics[width=0.9\linewidth]{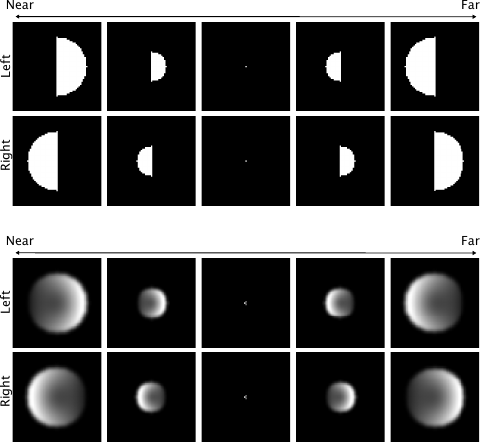}
    \caption{\textbf{Modelling parametric DP PSFs for realistic simulations}. (Top row) An ideal dual-pixel sensor would produce left, and right semi-circular PSFs. However, real-world DP PSFs look significantly different. The parametric model from  \cite{abuolaim2021learning} is used to simulate more realistic no-code DP PSFs (Bottom row).}
    \label{fig:dp-psf-modelling}
\end{figure}

For ours CADS framework, the coded DP PSFs are generated using the mask pattern and the naive (no code) DP PSFs. Modelling the left, right naive DP PSFs accurately is crucial for realistic simulations. The left, right dual-pixels receive light from different halves of the lens. Thus in an ideal scenario, the left, right naive DP PSFs are shaped as semi-circular kernels.  However, this is rarely seen in real-world DP PSFs. Errors in manufacturing, optical aberrations, and physical constraints for placement of microlenses and sensor well depths can cause light leakage from the other lens half~\cite{punnappurath2020modeling, abuolaim2021learning}, making it look more like tapered semi-circular halves. In \cite{abuolaim2021learning}, the authors designed a heuristic model to simulate DP PSFs that look closer to the real-world DP PSFs. The DP PSFs ($h^L_z$, $h^R_z$) are modeled as the Hadamard product of a 2D Butterworth filter with the circle-of-confusion. We generate our simulated naive DP PSFs in the same manner, choosing $n=1$ as the filter order, $\alpha=2.5$, $\beta=0.4$, and smoothing strength of 7. We generate PSFs for $N_z=21$ depth planes spanning equally both sides of the defocus. The left, right PSF z-stack corresponds to defocus blur sizes (or circle-of-confusion sizes) ranging from -40 to +40 pixels (signed blur size). Our no-code DP PSFs $h^L_z$, $h^R_z$ that were used in simulations are depicted in Fig.~\ref{fig:dp-psf-modelling}.
\subsection{Occlusion-aware dual-pixel image rendering}
For rendering accurate dual-pixel images, we adopt a multi-plane representation of the scene, where the scene is divided into discrete depth planes. Given the coded DP PSFs $h^{L,C}_z$ and $h^{R,C}_z$, the coded DP left, right images $I_L$, $I_R$ of a 3D scene can be expressed as a sum of 2D convolutions  
\begin{equation}
\begin{split}
    I_L = \sum_{k=0}^{K-1} h^{L,C}_{z_k} * s_{z_k}, \;\;\;\; I_R = \sum_{k=0}^{K-1} h^{R,C}_{z_k} * s_{z_k} ,
\end{split}\label{eq:imaging-model-simple}
\end{equation}
where $s_{z_k}$ is the scene intensity map which falls into the depth layer $k$ (which is at depth $z=z_{k}$), $K$ are the number of depth planes (or MPI planes), and $*$ is the 2D convolution operator. To remove artifacts at the edges of the MPI depth layers, we adopt the modifications to Eqn.~\ref{eq:imaging-model-simple} from Ikoma \textit{et al.}~\cite{ikoma2021depth}, thus we have a differentiable non-linear image formation model as follows
\begin{equation}
    \label{eq:ikoma-modification}
    \begin{split}
    I_L = \sum_{k=0}^{K-1} \frac{h^{L,C}_{z_k} * s_{z_k}}{E^L_k}\prod_{k'=k+1}^{K-1}(1-\frac{h^{L,C}_{z_{k'}} * \alpha_{z_{k'}}}{E^L_{k'}}), \\
    I_R = \sum_{k=0}^{K-1} \frac{h^{R,C}_{z_k} * s_{z_k}}{E^R_k}\prod_{k'=k+1}^{K-1}(1-\frac{h^{R,C}_{z_{k'}} * \alpha_{z_{k'}}}{E^R_{k'}}),
\end{split}
\end{equation}
where  $\alpha_{z_k}$ is the binary depth mask corresponding to depth layer $k$ (which is at depth $z=z_{k}$), $E^{L,R}_k$ are normalization factors equal to $h^{L,C}_{z_k} * \sum_{k'=0}^{k}\alpha_{k'}$ and $h^{R,C}_{z_k} * \sum_{k'=0}^{k}\alpha_{k'}$ respectively. We further add a minor modification to the above Eq~\ref{eq:ikoma-modification} to the binary depth masks (or alpha maps) $\alpha_k$. We first expand the alpha maps $\alpha_k$ into 3x3 2D max-pooling and then blend the maps into 2 adjacent depth layers instead of 1 (using 2x1x1 3D avg-pooling), and then normalize the alpha maps. Using $21$ depth planes is not a very coarse division of the scene into MPI layers, thus the blending still keeps the simulation realistic, while improving the rendering at edges of individual (and consecutive) MPI layers having fewer fringes/artifacts. 

\subsection{Evaluation Metrics}
We evaluate our simulation results and compare our proposed coded dual-pixel sensing approach (CADS) to naive DP and naive standard-pixel cases, as well as to previous works. We describe our evaluation metrics' definitions here \newline
\textbf{Depth Metrics.} For comparison with naive dual-pixel and naive standard-pixel, we use absolute metrics - 
\begin{itemize}
    \item RMSE (RMS Error): $\frac{1}{N}\sum_{i=1}^{N}|\hat{D}_i - D_i|^2$
    \item MAE (Mean Absolute Error): $\frac{1}{N}\sum_{i=1}^{N}|\hat{D}_i - D_i|$
    \item $\delta^1$ with threshold $T$: $\frac{1}{N}\sum_{i=1}^N\left(\text{max}\left(\frac{\hat{D}_i}{D_i}, \frac{D_i}{\hat{D}_i}\right) < T^1 \right)$
\end{itemize}
For comparison with previous works, we use affine-invariant metrics, as used in \cite{Garg_2019_ICCV}. 
\begin{itemize}
    \item AI(1): $\mathop{\min}\limits_{p,q}\left(\frac{\sum_{i=1}^{N}|D_i - (p\hat{D}_i+q)|}{N}\right)$
    \item AI(2): $\mathop{\min}\limits_{p,q}\left(\frac{\sum_{i=1}^{N}|D_i - (p\hat{D}_i+q)|^2}{N}\right)^{(1/2)}$
    \item $1-|\rho_s|$: $\rho_s$ denotes the Spearman's Rank Correlation Coefficient.  
\end{itemize}
\textbf{AIF Metrics.} For AIF deblurred predictions we compare results using PSNR, SSIM, and LPIPS~\cite{zhang2018unreasonable} metrics. 


\section{Results}
\subsection{End-to-end training results}
\begin{figure}[t!]
    \centering
    \includegraphics[width=\linewidth]{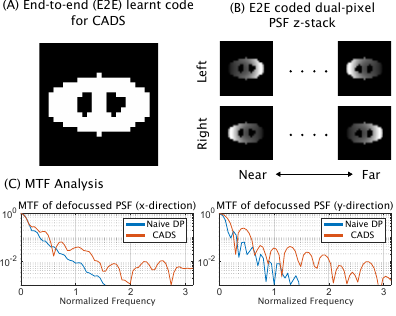}
    \caption{\textbf{CADS Learnt Mask.} (A) shows the learned amplitude mask and (B) shows the corresponding PSFs for different depths. (C) shows the MTF of CADS PSF vs. Naive DP PSF (at a defocussed depth, showing only left DP PSF). The higher MTF of the CADS PSF indicates better conditioning of the PSF.}
    \label{fig:e2e-result}
\end{figure}
We perform end-to-end training in simulations as outlined in the CADS Section (Sec. 3). We initialize the mask with a circular open aperture of pixelated size 21x21 (as mentioned in Sec 3.2.3). The mask resolution (in pixels) was chosen based on competing factors. A coarser pixelation (e.g., 7x7, 11x11) would mean the resulting blur would be less informative for small defocus blurs and perform worse. With finer pixelation one risks creating small feature sizes --- thereby causing severe diffraction-based distortion and having additional computational costs for end-to-end training --- hence the choice of 21x21. Different mask initializations were also tested; the circular open aperture initialization performed the best (on average, 0.35dB better in AIF and 4\% better in depth estimation). 

Thus, with the above-mentioned mask initialization, we train for 80 epochs on 20k FlyingThings3D scenes with a batch size of 8. The mask learning phase is only for the first 30 epochs, after which the mask is fixed. Cosine decay scheduling is applied for the learning of mask parameters ($\theta_C$) and CADNet weights ($\theta_D$) as well. In our mask parameterization, we initialize $\alpha$ to 0 and increase it using the schedule $\alpha_t = \alpha_0 + \frac{t}{8000}$, where $t$ is the total number of iterations. This is done to ensure smooth learning of a binary coded mask~\cite{shedligeri2017data}. An initial learning rate of $3\,\times\,10^{-5}$ and $3\,\times\,10^{-4}$ are used for CADNet weights ($\theta_D$) and for the mask parameters ($\theta_C$), alongwith cosine decay scheduling. During testing/inference, we explicitly threshold the mask, setting it to be binary. 

\subsection{End-to-end CADS Learnt Mask}
The learned mask and the corresponding left, right coded DP PSFs are shown in Fig.~\ref{fig:e2e-result}. The CADS PSFs are better conditioned as compared to the naive DP PSFs, as indicated by the MTF plots in Fig.~\ref{fig:e2e-result}(C). The learned coded aperture has the shape of a flattened ellipse (with a larger horizontal diameter), with two dots inside. The possible reason for this can be explained as follows. The intended goal is to learn a mask that gives the best defocus map prediction and AIF image prediction when coupled with dual-pixel sensors. In order to do so, the coded DP PSFs should be able to show a disparity effect with defocus, while having the shortest possible blur size. Since the disparity between the DP PSFs is horizontal, the learned coded aperture maintains the horizontal opening of the aperture, while reducing the vertical opening. Furthermore, the two opaque dots near the center on the horizontal axis potentially add to better conditioning of the mask and ensure recovery of high-frequency texture (mainly in the horizontal/x-direction). 
\subsection{Simulation results}
\subsubsection{Ablations on coded aperture}
\begin{figure}
    \centering
    \includegraphics[width=\linewidth]{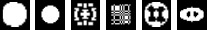}
    \caption{\textbf{Coded apertures considered for ablations}. We test out various coded apertures in conjunction with dual-pixel sensors in simulations. From left to right, showing no code mask, no code (50\% size) mask, Levin \etal~\cite{levin2007image} mask, separable MLS code~\cite{zhou2009coded}, Shedligeri \etal~\cite{shedligeri2017data} mask, and our proposed end-to-end learned CADS mask.}
    \label{fig:masks}
\end{figure}

Coded aperture masks have been used previously for PSF engineering to gain optimal deblurring and depth estimation performance~\cite{levin2007image, veeraraghavan2007dappled,asif2016flatcam,shedligeri2017data}. These have been used in the naive standard-pixel settings, and hence may not necessarily translate to optimal performance in the dual-pixel sensor setting. We compare the naive (no-code) DP case and our end-to-end learned code with the following codes in the dual-pixel sensor setting:  
\begin{itemize}
    \item Open aperture (no-code) that is 50\% smaller in area 
    \item Code from Levin \textit{et al.}~\cite{levin2007image}, which was derived based on an optimization problem formulated for constructing a desirable depth estimation mask. 
    \item Separable MLS code from \cite{asif2016flatcam} that has flat singular value spectrum and has been used for lensless imaging
    \item Code from Shedligeri \textit{et al.}~\cite{shedligeri2017data}. This code was learned end-to-end for the purposes of depth estimation, albeit in a standard pixel setting (not for dual-pixel sensors). 
\end{itemize}
The coded apertures are depicted in Fig.~\ref{fig:masks}. We trained CADNet-Mono for the above masks while keeping the mask fixed (no learning). The results of depth estimation and deblurring performance are given in Table~\ref{tab:coded-masks-ablations}. All the coded masks perform similarly in depth estimation and show gains in deblurring performance. Our proposed end-to-end learnt code possesses the novelty of being the first one to be trained specifically for dual-pixel sensor imaging. Hence, our end-to-end learnt code outperforms all the above coded masks to give the best depth estimation and deblurring performance.  
\begin{table}[t!]
    \centering
    \begin{tabular}{l|c|c|c}
        \multicolumn{1}{c|}{CADS Mask} & \multicolumn{1}{c|}{Depth Pred.} &
        \multicolumn{2}{c}{AIF Pred.} \\ \hline
          & MAE(mm) $\downarrow$ & PSNR $\uparrow$ & SSIM$\uparrow$ \\ \hline
        No code  &  5.51 &                      29.7 & 0.83\\
No code (50\%)      &                    5.77 & \cellcolor{tabsecond}30.6 & \cellcolor{tabsecond}0.85 \\
Levin \textit{et al.}~\cite{levin2007image}    &                      5.54 &  \cellcolor{tabsecond}30.6 & \cellcolor{tabsecond}0.85 \\
MLS code~\cite{asif2016flatcam}     &                      5.64 &                      30.3 & \cellcolor{tabsecond}0.85 \\
Shedligeri \textit{et al.}~\cite{shedligeri2017data}     & \cellcolor{tabsecond}5.49 &                      30.4 & \cellcolor{tabsecond}0.85 \\ \hline 
E2E learnt (ours)  &  \cellcolor{tabfirst}5.15 &  \cellcolor{tabfirst}31.2 & \cellcolor{tabfirst}0.87
    \end{tabular}
    \caption{\textbf{Coded Mask Ablation.} Coded aperture DP outperforms naive DP for AIF and depth prediction. Among coded DP designs, the proposed end-to-end learned design offers the best performance. Red indicates best, orange indicates second best.}
    \label{tab:coded-masks-ablations}
\end{table}

\subsubsection{Comparison with existing DP-sensing works}
\begin{table*}[ht!]
    \centering
    \begin{tabular}{c|c|c|c|c|c} 
    FlyingThings3D dataset & \multicolumn{2}{c|}{AIF Predictions} & \multicolumn{3}{c}{Disparity Predictions} \\ \hline 
    Method & PSNR(dB)$\uparrow$ & SSIM$\uparrow$ & AI(1)$\downarrow$ & AI(2)$\downarrow$ & $1 - |\rho_s|$ $\downarrow$ \\ \hline
 DPDNet~\cite{abuolaim2021learning}& 24.34 & 0.63 & N.A. & N.A. & N.A. \\ 
         DDDNet~\cite{pan2021dual}&21.35& 0.54 & 0.277 & 0.381 & 0.561 \\   
         Xin \etal ~\cite{xin2021defocus}$^{\dagger}$ & 18.13 & 0.30
      & 0.302 & 0.415 & 0.951 \\   
 Punnappurath \etal ~\cite{punnappurath2020modeling}$^{\dagger}$ & N.A. & N.A.& 0.194 & 0.264 & 0.243 \\ 
 Kim \etal~\cite{Kim_2023_CVPR} & N.A. & N.A.& 0.287 & 0.382 & 0.694 \\ \hline
 Naive DP (ours) & \cellcolor{tabsecond}29.72 & \cellcolor{tabsecond}0.83 & \cellcolor{tabsecond}0.019 & \cellcolor{tabsecond}0.050 & \cellcolor{tabsecond}0.092 \\
 \textbf{CADS (ours)} & \cellcolor{tabfirst}31.20 & \cellcolor{tabfirst}0.87 & \cellcolor{tabfirst}0.018 & \cellcolor{tabfirst}0.046 & \cellcolor{tabfirst}0.087 \\\hline\hline 
 \multicolumn{1}{l|}{NYUv2 dataset} & \multicolumn{5}{c}{} \\ \hline
  DPDNet~\cite{abuolaim2021learning}& 26.90 & 0.80 & N.A. & N.A. & N.A. \\ 
         DDDNet~\cite{pan2021dual}& 20.42 & 0.56 & 0.303 & 0.392 & 0.595 \\   
         Xin \etal ~\cite{xin2021defocus}$^{\dagger}$ & 16.70 & 0.37
      & 0.302 & 0.412 & 0.579 \\   
 Punnappurath \etal ~\cite{punnappurath2020modeling}$^{\dagger}$ & N.A. & N.A. & 0.149 & 0.204 & 0.170 \\ 
 Kim \etal~\cite{Kim_2023_CVPR} & N.A. & N.A.& 0.306 & 0.386 & 0.489 \\ \hline
 Naive DP (ours) & \cellcolor{tabsecond}29.33 & \cellcolor{tabsecond}0.87 & \cellcolor{tabsecond}0.017 & \cellcolor{tabsecond}0.030 & \cellcolor{tabsecond}0.058 \\
 \textbf{CADS (ours)} & \cellcolor{tabfirst}31.32 & \cellcolor{tabfirst}0.91 & \cellcolor{tabfirst}0.016 & \cellcolor{tabfirst}0.029 & \cellcolor{tabfirst}0.057
    \end{tabular}
    \caption{\textbf{Comparison with existing DP methods.} CADS offers the best AIF and disparity estimation quality on our simulated DP captures based on FlyingThings3D scenes, and on our simulated DP captures based on the NYUv2 scenes. Red highlights best, orange highlights second best. $^{\dagger}$ indicates metrics computed over 16 samples since these methods had a slow runtime. For methods where AIF/disparity is not predicted, metrics are marked as N.A.}
    \label{tab:compare-with-previous-suppl}
    \vspace{-10pt}
\end{table*}
We test existing DP-sensing works on simulated naive DP captures based on FlyingThings3D scenes and also on simulated naive DP captures based on NYUv2 scenes. Since existing methods were designed for reconstructing from naive DP captures, we created a simulated dataset of naive (no code) DP captures, using the FlyingThings3D dataset scenes and another one using NYUv2 dataset scenes. We compare the following works
\begin{itemize}
    \item \textbf{DPDNet} - the authors in \cite{abuolaim2020defocus} designed a UNet-based neural network to reconstruct the deblurred (all-in-focus) image from DP captures, called DPDNet. DPDNet was trained using supervised real-world AIF GT data. We use the same trained model weights as given in their \hyperlink{https://github.com/Abdullah-Abuolaim/defocus-deblurring-dual-pixel}{repository}.
    \item \textbf{DDDNet} - the authors in \cite{pan2021dual} designed a two-stage neural network (called DDDNet) to predict the disparity map and the deblurred (all-in-focus) image of the scene from DP captures. We use the same model weights as given in their \hyperlink{https://github.com/panpanfei/Dual-Pixel-Exploration-Simultaneous-Depth-Estimation-and-Image-Restoration}{repository}. 
    \item \textbf{Xin \etal 2021}. In \cite{xin2021defocus}, an optimization problem was formulated for simultaneous defocus map and all-in-focus image recovery. We use the code given in their \hyperlink{https://github.com/cmu-ci-lab/dual_pixel_defocus_estimation_deblurring}{repository} and pass our simulated DP PSFs as arguments for the optimization problem. Owing to the fact that the runtime for this was slow, we evaluate for a randomly selected set of 16 DP captures. 
    \item \textbf{Punnappurath \etal 2020}. In \cite{punnappurath2020modeling}, an optimization problem was formulated for recovering the disparity map from DP captures, exploiting left-right kernel symmetry to do so. We use the code given in their \hyperlink{https://github.com/abhijithpunnappurath/dual-pixel-defocus-disparity}{repository}. Due to slow inference time, we evaluate for a randomly selected set of 16 DP captures.
    \item \textbf{Kim \etal 2023}. In \cite{Kim_2023_CVPR}, the authors trained a stereo disparity estimation network to handle bi-directional disparity, and then devised a self-supervised loss to learn disparity based on the DPDBlur~\cite{abuolaim2020defocus} dataset. We evaluate the code given on their \hyperlink{https://github.com/KAIST-VCLAB/dual-pixel-disparity}{repository}. 
\end{itemize}
Table~\ref{tab:compare-with-previous-suppl} shows the quantitative results of previous methods with our CADS method, and with our method for the naive DP case as well. The simulated DP captures showed bidirectional disparity in the left, right captures. Thus, methods designed for uni-directional disparity did not work well~\cite{pan2021dual, xin2021defocus}. Punnappurath \etal~\cite{punnappurath2020modeling} outputs disparity maps that resemble the ground truth but are not as accurate as our methods. While Kim \etal~\cite{Kim_2023_CVPR} is trained to handle bi-directional disparity, the error is higher, possibly due to the fact that the self-supervised loss is not trained on the simulated captures.  

\subsection{DSLR Photography results}
\subsubsection{Experimental setup}
\begin{figure}[h!]
    \centering
    \includegraphics[width=\columnwidth]{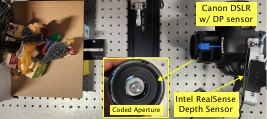}
    \caption{\textbf{Experimental setup for DSLR photography captures.} The coded aperture is placed in the aperture plane inside the Yongnuo lens.}
    \label{fig:dslr-setup}
\end{figure}

For real-world DSLR photography experiments, we use the Canon EOS 5D Mark IV DSLR. It is equipped with a 30MP color sensor (6880x4544 pixels) with a R-G-G-B Bayer pattern, with pixel pitch $p=5.36~\mu$m. For DSLR photography, we capture naive DP images and CADS images with a Yongnuo $50$~mm focal length lens with aperture $L=f/4=12.5$~mm. We print our binary amplitude mask having $12.5$~mm diameter on transparency sheets, and place it inside the DSLR lens, as done in \cite{levin2007image}. Fig.~\ref{fig:dslr-setup} illustrates the same. We set the focus distance of the camera to $40$~cm, and we set up toy scenes $32$--$53$~cm away from the camera. These imaging parameters of the setup were chosen such that the defocus blur sizes will approximately be within 0-40 pixel size (on both sides of the defocus).

\textbf{Canon Dual-pixel RAW data capture}. We capture images using the Canon camera, under the DP-RAW setting with the lowest possible ISO setting of $100$. We process the raw .CR2 files using \hyperlink{https://helpx.adobe.com/camera-raw/using/adobe-dng-converter.html}{Adobe DNG converter} to convert them to the DNG format. We extract the combined (L+R) and left (L) images from the DNG files using the Tiff() capabilities in MATLAB. For simplicity we do not perform demosaicking, thus we obtain RGB DP captures of size 3440x2272x3. The 14-bit RAW data is appropriately scaled and the black-level is subtracted to get the left, right DP images. 

\begin{figure*}
    \centering
    \includegraphics[width=\linewidth]{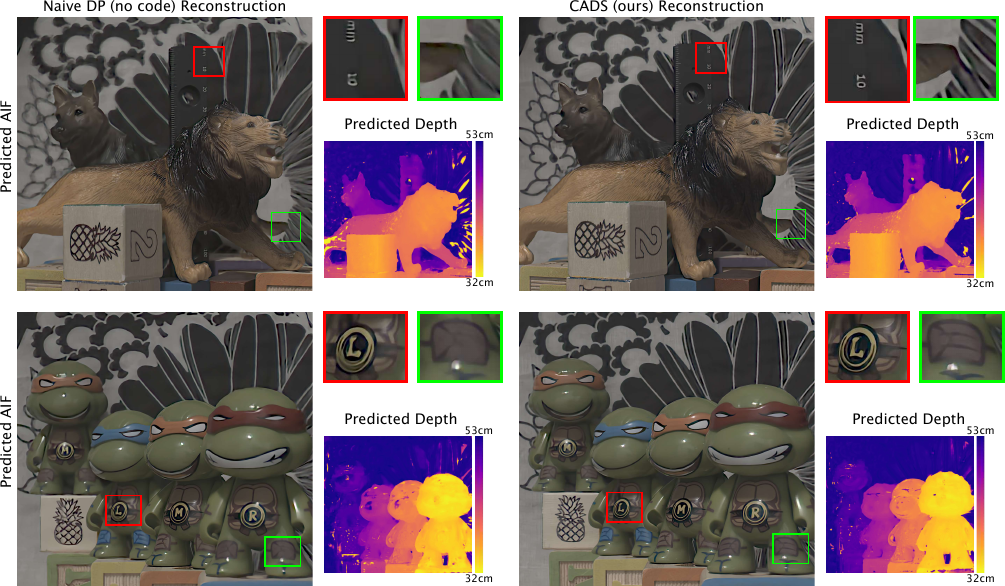}
    \caption{\textbf{DSLR photography additional results}. Showing Naive DP and CADS (ours) reconstructions (depth and AIF). Zoom-in for better comparison.}
    \label{fig:macro-res-suppl}
\end{figure*}

\textbf{Scene capture details}. For a given scene, we capture a naive (no code) DP measurement, a CADS measurement, and also a $f/22$ measurement to obtain the ground truth deblurred all-in-focus image of the scene. Furthermore, we also capture a coarse ground truth depth map using an Intel RealSense D415 stereo sensor (see Fig.~\ref{fig:dslr-setup}). We pre-calibrate the Intel RealSensor sensor with our Canon DSLR sensor with the help of a 10x12 checkerboard pattern, so as to transform the depth map into the DSLR's frame of reference. We crop the captured DP measurements and only reconstruct the central 1696x1522x3 region. We show a few more example results in Fig.~\ref{fig:macro-res-suppl}.  

\subsubsection{PSF capture, calibration, and fine-tuning}
\begin{figure}[h!]
    \centering
    \includegraphics[width=\linewidth]{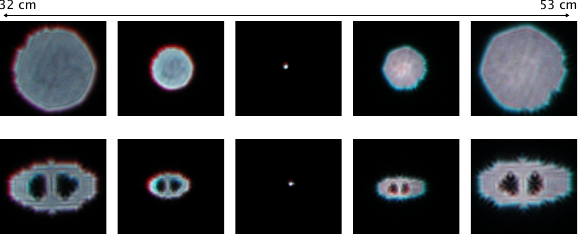}
    \caption{\textbf{DSLR PSF captures}. Showing combined (L+R)  real-world PSFs captured at different depths. (Top row) no-code case. (Bottom row) CADS case.}
    \label{fig:dslr-psf-captures}
\end{figure}

\textbf{Real-world PSF capture}. To capture PSFs, we capture pinhole images (illuminated with white light) for 21 depths ranging from $32$~cm to $53$~cm. We capture these for the no-code (open aperture) case and for the E2E learnt CADS case. Fig.~\ref{fig:dslr-psf-captures} illustrate the same. 

\textbf{Fine-tuning}. To reconstruct real-world captures, we perform fine-tuning of a trained CADNet-RGB model (trained on simulated FlyingThings3D scenes using simulated DP PSFs). We first capture real-world PSFs as mentioned above. The real-world PSFs are used to simulate DP captures based on FlyingThings3D scenes, and CADNet-RGB model weights are fine-tuned on the new captures for 30 epochs. During fine-tuning phase, we train for a variable amount of heteroscedastic noise~\cite{foi2008practical} levels ranging from $0.7\%$--$1.5\%$, along with extra data augmentations (random) on brightness, contrast, gamma, and hue. This is done to remove certain sim-to-real mismatches to enable better depth and AIF reconstructions. 

\textbf{Calibration}. In \cite{xin2021defocus}, the authors describe a vignetting calibration scheme. Images of a white sheet are captured to model the vignetting profile for left, right DP images. Scenes are pre-processed by dividing the left, right scene captures with the corresponding vignetting images. We follow the same procedure in our case as well. 

\subsection{Endoscopy, Dermoscopy results}
\subsubsection{Experimental setup details}
\begin{figure}
    \centering
    \includegraphics[width=\linewidth]{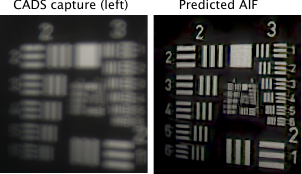}
    \caption{\textbf{USAF reconstruction result}. With our CADS prototype endoscope, we are able to resolve $\leq40~\mu$m features, as line pairs in Group (3,5) are visible in the reconstruction. USAF target placed $\sim$10mm away from scope.}
    \label{fig:usaf-suppl}
\end{figure}

\begin{figure}[h!]
    \centering
    \includegraphics[width=\columnwidth]{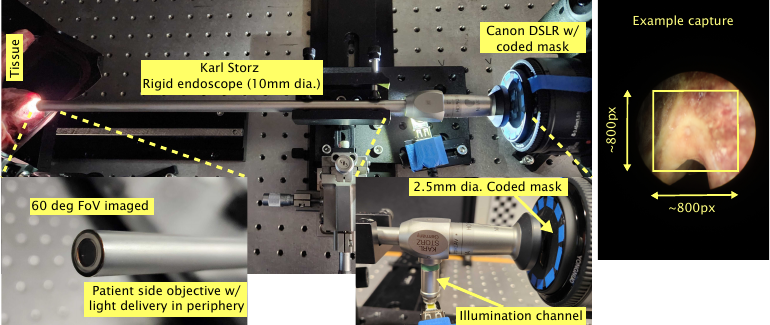}
    \caption{\textbf{CADS Endoscopy setup}. (Left) CADS endoscope setup, with zoom-in insets. (Right) Showing an example RAW capture.}
    \label{fig:endoscopy-setup-suppl}
\end{figure}

\textbf{Endoscopy setup and capture details}. We use a Karl Storz 26003ARA Rubina rigid endoscope of 10~mm diameter. A rigid endoscope consists of several relay lenses to relay the image from the patient side to the surgeon side (eyepiece). For making a prototype CADS endoscope, we mount a 2.5mm coded mask in front of the Canon DSLR lens and align its optical axis to that of the endoscope. We focus the DSLR such that it is focussed at a point 20~mm away from the other end of the endoscope. Fig.~\ref{fig:endoscopy-setup-suppl} illustrates ours CADS endoscope prototype. We capture a 2D PSF array 2.5~mm away from the scope, and scale it down to obtain PSFs at 21 depths ranging from 2.5~mm to 20~mm. We perform fine-tuning as outlined before but with some more modifications -- (1) we implicitly model PSF spatial variance across the FoV (see ahead for more details), (2) we only reconstruct for negative defocus, and (3) along with data augmentations, we add random bias and random attenuation to one of the channels (because the vignetting correction is not perfect). With such a setup, we are able to capture $\leq40~\mu$m features over an extended depth-of-field, as illustrated in Fig.~\ref{fig:usaf-suppl}. 

\textbf{Dermoscopy setup and capture details}. We built a prototype CADS dermoscope using the Pixel 4 camera, 12x macro lens, and a 2.5~mm diameter CADS mask. We focus at the closest distance possible i.e. 45~mm. We capture PSFs at 32~mm and scale them accordingly to obtain PSFs at 21 depth planes from 32~mm to 76~mm. We perform fine-tuning in the same way as outlined for the endoscopy case. We capture DP data using an open-source Android app \url{https://github.com/google-research/google-research/tree/master/dual_pixels}. Since the data obtained has rectangular-shaped pixels (2:1) we re-size the smaller dimension accordingly and reconstruct for a central FoV of 512x432 pixels. 

\subsubsection{Modelling spatially-varying PSFs}
\begin{figure}[h!]
    \centering
    \includegraphics[width=\linewidth]{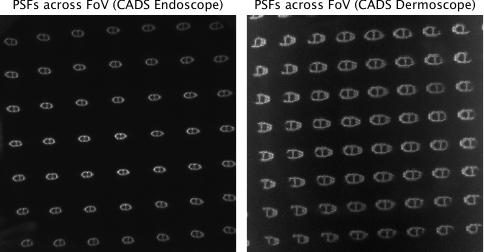}
    \caption{\textbf{Spatial variance of PSFs across FoV}. Showing for endoscopy and dermoscopy cases (left and right respectively).}
    \label{fig:svpsf}
\end{figure}

\begin{figure}[h!]
    \centering
    \includegraphics[width=\columnwidth]{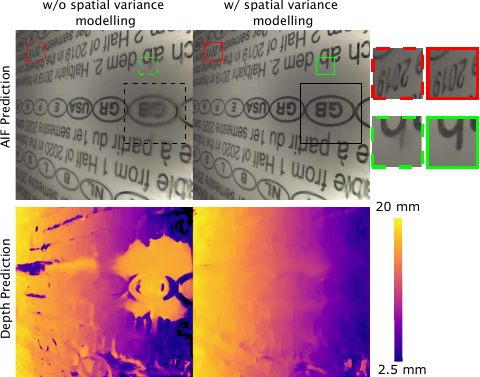}
    \caption{\textbf{Modelling spatial variance in PSFs across FoV.} For coded dual-pixel endoscopy captures, we observe that accounting for several PSFs across the FoV during fine-tuning phase enables better reconstruction in AIF and especially in depth predictions.}
    \label{fig:endoscopy-spv-modelling}
\end{figure}

For our CADS endoscopy and dermoscopy setup, we observe that the coded DP PSFs vary over the entire field of view (FoV), as shown in Fig.~\ref{fig:svpsf}. This leads to improper reconstruction results if we fine-tune CADNet with the central PSF only. To account for this spatial variance, we capture a 2D array of PSFs across the FoV. During the fine-tuning phase, we randomly sample one of the PSFs (out of many) in every iteration and use its corresponding PSF z-stack to simulate and render out the coded dual-pixel images. By doing so, the CADNet network sees all the variations in the PSF for the given system and thus can correct for those reliably. Fig.~\ref{fig:endoscopy-spv-modelling} illustrates the same. For the case in which we fine-tune only with the central PSF z-stack, we obtain incorrect depth maps and slightly incorrect AIF maps. Our fine-tuning using the captured PSF 2D array gives better AIF reconstruction and much better depth reconstruction. We use the same fine-tuning procedure (using a captured PSF 2D array) for our dermoscopy experiments. 
